\documentclass[aip,reprint,graphicx]{revtex4-1}
\usepackage{graphicx} 
\usepackage{epstopdf} % converts eps->pdf so that article can include both the eps and the pdf figures
\usepackage{amsmath}
\usepackage{amsfonts}
\usepackage{amssymb}
\usepackage{bm}
\usepackage{fixme} % \fxnote*{question}{text}
\fxsetup{
	status=draft,
	author={Note},
	layout=footnote, % inline, footnote or pdfnote
	theme=color
}

\newcommand{\ci}{\mathrm{i}}
\newcommand{\vekn}[1]{\boldsymbol{\mathrm{#1}}}  	% Bold face and non-italized
\newcommand{\iim}{\mathrm{i}}	% imaginary symbol
\newcommand{\ud}{\mathrm{d}}	% differantial d 
\newcommand{\kommentti}[1]{}
\newcommand{\ii}{\mathrm{i}}
\newcommand{\dd}{\mathrm{d}}

\begin{document}
%\setboolean{shortarticle}{false} % true = letter, false = research article

\title{Modeling open nanophotonic systems using the Fourier modal method: Generalization to 3D Cartesian coordinates}

\affiliation{DTU Fotonik, Department of Photonics Engineering, Technical University of Denmark, \O rsteds Plads, Building 343, DK-2800 Kongens Lyngby, Denmark}

\author{Teppo H\"{a}yrynen}
\author{Andreas Dyhl Osterkryger}
\author{Jakob Rosenkrantz de Lasson}
\author{Niels Gregersen}
\email{ngre@fotonik.dtu.dk}

\date{\today}


%\ociscodes{
%(050.1755) Computational electromagnetic methods; 
%(350.3950) Micro-optics;
%(230.7370) Waveguides
%(000.3860) Mathematical methods in physics; 
%(000.4430) Numerical approximation and analysis; 
%}

%\doi{\url{http://dx.doi.org/10.1364/josaa.XX.XXXXXX}}

% Packages
%\usepackage[dvips]{graphicx} % ps-figures
%\usepackage[english]{babel}

%\graphicspath{{figures/}}

\begin{abstract}
Recently, an open geometry Fourier modal method based on a new combination of an open boundary condition and a non-uniform $k$-space discretization was introduced for rotationally symmetric structures providing a more efficient approach for modeling nanowires and micropillar cavities [J. Opt. Soc. Am. A 33, 1298 (2016)]. Here, we generalize the approach to three-dimensional (3D) Cartesian coordinates allowing for the modeling of rectangular geometries in open space.
The open boundary condition is a consequence of having an infinite computational domain described using basis functions that expand the whole space. The strength of the method lies in discretizing the Fourier integrals using a non-uniform circular "dartboard" sampling of the Fourier $k$ space. We show that our sampling technique leads to a more accurate description of the continuum of the radiation modes that leak out from the structure. We also compare our approach to conventional discretization with direct and inverse factorization rules commonly used in established Fourier modal methods. We apply our method to a variety of optical waveguide structures and demonstrate that the method leads to a significantly improved convergence enabling more accurate and efficient modeling of open 3D nanophotonic structures.
\end{abstract}

%\setboolean{displaycopyright}{true}

\maketitle
%\thispagestyle{fancy}
%\ifthenelse{\boolean{shortarticle}}{\abscontent}{}

%\listoffixmes

\section{Introduction}

Numerous nanophotonic devices including microcavity resonators \cite{Vahala2003}, slow-light waveguides \cite{Lecamp2007b, MangaRao2007, Mork2009} and single-photon sources \cite{Gregersen2013, Aharonovich2016} are open systems with properties strongly characterized by their leakage of light into the surroundings, that in principle extend to infinity. With the exception of the Green's function integral equation approach \cite{Lavrinenko2014}, the majority of conventional approaches for modeling photonic nanostructures including the finite-difference time-domain technique \cite{Taflove2004,Lavrinenko2014} and the finite-elements method \cite{Lavrinenko2014} inherently rely on a limited computational domain combined with either periodic, closed or artificially absorbing boundary conditions (BCs). That is, most conventional methods cannot fully account for the openness of a system, although this is required to correctly model radiative losses. Therefore, simulations of open systems require careful treatment of the boundaries of the computational domain to avoid artificial reflections from the domain wall \cite{Berenger1994, Hugonin2005a, Gregersen2008a, Gregersen2010}. 

To circumvent the problem of selecting a proper artificial absorbing BC \cite{DeLasson2015a}, we have developed a Fourier modal method based on a new combination of an open BC and an efficient discretization scheme \cite{Hayrynen2016}, called oFMM in the following. The formalism presented in \cite{Hayrynen2016} was, however, limited to rotationally symmetric structures. In this work, we apply both the open boundary formalism and the efficient sampling of the $k$ space to model general 3D structures in Cartesian coordinates, in particular the rectangular waveguide. We remark that the new oFMM formalism affects only the eigenmode calculation; when they have been computed, the oFMM formalism is otherwise identical to the well-established Fourier modal method.
 
The open boundary of the computational domain can be described by using basis functions, plane waves in this case, expanding the whole infinite space and by using the Fourier transformation. This formalism replaces the usual Fourier series expansion \cite{Li1997, Li1996b, Lalanne1998}, which inherently assumes periodicity of the field components.  
While the use of the Fourier integral transformation gives an exact description of the structure in the limit of continuous $k$-space sampling, the numerical implementation does require a $k$-space discretization. An advantage of the new approach, however, is that we have the freedom to choose the $k$-space discretization in a way that leads to a more efficient mode sampling. 
Similar ideas have also been reported for two-dimensional (2D)~\cite{Guizal2003} and rotationally symmetric three-dimensional (3D)~\cite{Bonod2005, Bava2001, Dems2010} structures, but without applying efficient $k$-space discretization schemes. In contrast, in our recent work \cite{Hayrynen2016} we developed the oFMM approach based on open BCs and a Chebyshev grid \cite{Boyd2001,DeLasson2012} for rotationally symmetric structures, an approach which we here generalize for any 3D system. 

In addition, we discuss how to use Li's factorization rules \cite{Li1997, Li1996b, Lalanne1998} in connection with the 3D oFMM method. It turns out that, while in the rotationally symmetric case Li's factorization rules are straightforwardly adopted for any $k$-space discretization \cite{Bonod2005, Hayrynen2016}, for the general 3D approach we can only apply the inverse factorization rule when using the conventional discretization scheme. In spite of this subtlety, we will show that our new discretization scheme leads to a faster convergence compared to traditional discretization schemes.

The manuscript is organized as follows. Section \ref{sec:theory} outlines the theory of the oFMM approach. The details of the new discretization scheme are discussed in Section \ref{sec:disc}. The method is tested by calculating the dipole emission in a waveguide and the reflection of the fundamental mode from a waveguide-metal interface in Section \ref{sec:results}. After a discussion of advantages and limitations of the method in Section \ref{sec:discussion}, conclusions are drawn in Section \ref{sec:conc}, and detailed derivations of our theory are provided in the Appendix.

\section{Theory} \label{sec:theory}

In this section, we follow the approach of Ref. \cite{Hayrynen2016} and generalize the results for the 3D Cartesian coordinate system. We outline the derivation of the open BC formalism and introduce the theoretical concepts required to understand the results of the following sections. As important examples we show how the oFMM approach is applied to calculate the emission from a dipole placed inside a waveguide and to compute the reflection from a waveguide-metal interface. In Appendix \ref{app:deriv}, we give the detailed derivations of the open geometry formalism and discuss the applicability of the Fourier factorization rules.

\subsection{Open boundary condition formalism}

We use a complete vectorial description of Maxwell's equations based on Fourier expansion and open BCs to describe the electro-magnetic (EM) fields in a $z$-invariant material section. The $z$ dependence can be treated by combining $z$-invariant sections using the scattering matrix formalism (see, e.g. \cite{Li1996a,Lavrinenko2014} for details); this part of the calculation remains unchanged by the new oFMM formalism which only alters the way that modes of each $z$-invariant section are calculated.
The task is then to compute the lateral electric and magnetic field components of the eigenmodes, which form the expansion basis for the EM field.
In the conventional FMM, this is done by expanding the field components as well as the permittivity profile in Fourier \textit{series} in the lateral coordinates $(x,y)$ on a \textit{finite-sized} computational domain, implying that these functions vary \textit{periodically} in these coordinates. In the open boundary formalism, we instead consider an \textit{infinite-sized} computational domain and employ expansions in Fourier \textit{integrals}. We use a plane-wave expansion as basis functions. In the following, we describe the general steps and equations required to expand the field components and to solve for the expansion coefficients and the propagation constant. The specific equations and derivations are given in Appendix \ref{app:deriv} and referenced throughout this section.    

We start by considering a $z$-invariant part of the space where the lateral structure is defined by the relative permittivity $\varepsilon(x,y)$ and impermittivity $\eta(x,y) \equiv 1/\varepsilon(x,y)$. For simplicity, we consider a non-magnetic material having vacuum permeability $\mu_0$. In such a region of space, we write the Maxwell's equations using a harmonic time dependence of the form $\exp \left(- \ci \omega t\right)$ as
\begin{eqnarray}
\nabla \times \mathbf{E}(x,y,z) &=& \iim\omega\mu_0 \mathbf{H}(x,y,z) \label{eq:ME1} \\
\nabla \times \mathbf{H}(x,y,z) &=& -\iim\omega \varepsilon_0 \varepsilon(x,y) \mathbf{E}(x,y,z), \label{eq:ME2}
%\label{eq:}
\end{eqnarray}
where $\omega$ is the angular frequency and $\mathbf{E}$ and $\mathbf{H}$ are the vectorial electric and magnetic fields respectively. We then write the fields in a component-wise representation as shown in Eqs. (\ref{eqn:MaxHx})-(\ref{eqn:MaxEz}) in Appendix \ref{app:deriv} and introduce a $z$ dependence of the form $\exp\left(\pm \ci \beta z\right)$, where $\beta$ is the propagation constant of a particular eigenmode. The individual field components and the permittivity and impermittivity functions are then expanded on plane wave basis functions $g(k_x, k_y) = \exp[\ii (k_x x + k_y y) ]$ as
\begin{eqnarray}
\nonumber
&& f(x,y) \\
\nonumber
&=& \int_{-\infty}^{\infty} \int_{-\infty}^{\infty} c_f(k_x,k_y) g(k_x, k_y) \ud k_x \ud k_y \\
&\simeq& \sum_m \sum_l c_f(k_x^m,k_y^l) g(k_x^m, k_y^l) \Delta k_x^m \Delta k_y^l \label{eq:FourExpa1}
\end{eqnarray}
where in the last row the integral expansions are discretized using a Riemann sum on a $(k_x^m,k_y^l)$ grid for numerical calculations. The double summation over the indices $m$ and $l$ (\ref{eq:FourExpa1}) is valid for the conventional \textit{separable} discretization scheme, where the discretization grid coordinates along the $k_x$ and $k_y$ axes in $k$ space are defined \textit{independently} of each other.
However, when using a non-separable representation as we will do in the following, the Riemann sum is instead written as 
\begin{equation}
f(x,y) \simeq \sum_\xi c_f(k_x^\xi,k_y^\xi) g(k_x^\xi, k_y^\xi) \Delta k_\xi, \label{eq:FourExpa2}
\end{equation}
where a single index $\xi$ is used to describe the discretization points in the 2D $k$ space and $\Delta k_{\xi}$ is the discretization area for the $\xi$'th $k$ point. In the particular case of the separable discretization in (\ref{eq:FourExpa1}), we have $\Delta k_{\xi} = \Delta k_x^m \Delta k_y^l$. This discretization area $\Delta k_\xi$ will generally vary as function of $\xi$. Furthermore, as will be discussed in detail in Section \ref{sec:disc}, the selection of the wave number values $k_x^\xi$ and $k_y^\xi$ defines the Fourier expansion basis. The computational efficiency of our approach depends crucially on the choice of this expansion basis.  

When expanding the field components using the separable discretization, we treat the product of the permittivity function and the electric field components in Eq. (\ref{eq:ME2}) using Li's factorization rules \cite{Li1997, Li1996b, Lalanne1998}. However, as discussed in Appendix \ref{app:deriv}, this is not possible when using a non-separable discretization. The details of the expansions are given in Eqs. (\ref{eq:drf})-(\ref{eq:drff}), (\ref{eq:ToepExp})-(\ref{eq:ezd}) and (\ref{eq:ifdiv})-(\ref{eqn:EpsY}).
After inserting the expansions into Maxwell's equations and eliminating the $z$ components of the EM fields, we arrive at two set of equations that couple the lateral field components (Eqs. (\ref{eq:ExEyMat}) and (\ref{eq:HxHyMat}) in Appendix \ref{app:deriv})
\begin{align}
\left[ \begin{matrix}
\mathbf{k}_x \bm{\varepsilon}_{\mathrm{Tot}}^{-1} \mathbf{k}_y & -\mathbf{k}_x \bm{\varepsilon}_{\mathrm{Tot}}^{-1} \mathbf{k}_x + k_0^2 \mathbf{I} \\
\mathbf{k}_y \bm{\varepsilon}_{\mathrm{Tot}}^{-1} \mathbf{k}_y - k_0^2 \mathbf{I} & -\mathbf{k}_y \bm{\varepsilon}_{\mathrm{Tot}}^{-1} \mathbf{k}_x
\end{matrix} \right] \left[ \begin{matrix}
\mathbf{h}_x\\
\mathbf{h}_y
\end{matrix} \right] &= \pm \omega \varepsilon_0 \beta \left[ \begin{matrix}
\mathbf{e}_x \\
\mathbf{e}_y
\end{matrix}\right]  \label{eq:ecomp} \\
\left[ \begin{matrix}
-\mathbf{k}_x \mathbf{k}_y & \mathbf{k}_x^2 - k_0^2 \bm{\varepsilon_y} \\
k_0^2 \bm{\varepsilon_x} - \mathbf{k}_y^2 & \mathbf{k}_y \mathbf{k}_x
\end{matrix} \right] \left[ \begin{matrix}
\mathbf{e}_x \\
\mathbf{e}_y
\end{matrix} \right] &= \pm \omega \mu_0 \beta \left[ \begin{matrix}
\mathbf{h}_x \\
\mathbf{h}_y
\end{matrix} \right], \label{eq:hcomp}
\end{align}
where $\mathbf{e}_x$, $\mathbf{e}_y$, $\mathbf{h}_x$ and $\mathbf{h}_y$ are the vectors of the expansion coefficients of $E_x$, $E_y$, $H_x$, and $H_y$, respectively, and $\mathbf{k}_x$ and $\mathbf{k}_y$ are diagonal matrices of the discretized $k_x^\xi$ and $k_y^\xi$ values. 
Furthermore, $\bm{\varepsilon}_{\mathrm{Tot}} =  \bm{\Delta\varepsilon} \bm{\Delta k} + \varepsilon_B \mathbf{I}$ with $\bm{\Delta\varepsilon}$ being the Toeplitz matrix defined in Eq. (\ref{eq:epsmdef}), $\mathbf{I}$ is the identity operator and $\bm{\Delta k}$ is the diagonal matrix containing the elements $\Delta k_\xi$. When using a separable discretization grid, $\bm{\varepsilon}_x$ and $\bm{\varepsilon}_y$ are given by Eqs. (\ref{eqn:EpsX}) and (\ref{eqn:EpsY}) respectively. 
Combining Eqs. (\ref{eq:ecomp}) and (\ref{eq:hcomp}) allows us to compute, for example, the lateral electric field components $E_{x,j}(x,y)$ and $E_{y,j}(x,y)$ of the eigenmode $j$ and its propagation constant $\beta_j$, after which the lateral magnetic field components $H_{x,j}(x,y)$ and $H_{y,j}(x,y)$ and the longitudinal field components $E_{z,j}(x,y)$ and $H_{z,j}(x,y)$ can be derived.   
In Appendix \ref{app:deriv}, we show how Li's factorization rules are correctly used with the oFMM based on equidistant discretization. However, our non-separable "dartboard" discretization introduced in Section \ref{sec:disc} is not compatible with the inverse factorization rule, and for this reason we employ only the direct factorization rule, which means that we use $\bm{\varepsilon}_x= \bm{\varepsilon}_y= \bm{\varepsilon}_{\mathrm{Tot}}$. 

\subsection{Field emitted by a point dipole} \label{sec:emission}

In the modal expansion method, the emission from a point dipole placed in a photonic structure can be described \cite{Lavrinenko2014} as an expansion of eigenmodes with expansion coefficients proportional to the electric field strength of the corresponding eigenmode obtained from Eqs. (\ref{eq:ecomp})-(\ref{eq:hcomp}) at the emitter position. The total field emitted by a point dipole \textbf{p} placed at $\mathbf{r}_{\mathrm{pd}}$ inside a $z$-invariant structure can be represented as
\begin{eqnarray}
\nonumber
&&\mathbf{E}(x,y,z) = \sum_{j}a_j(\mathbf{r}_{\mathrm{pd}},\mathbf{p}) \mathbf{E}_j(x,y,z) \\ 
&=& \sum_{j}\sum_{\xi} a_j(\mathbf{r}_{\mathrm{pd}},\mathbf{p}) c_{j,\xi} \mathbf{g}_{\xi}(x,y)\Delta k_{\xi}e^{\pm\iim\beta_j (z-z_{\mathrm{pd}})},
\label{eq:em_Ef}
\end{eqnarray}
where $a_j(\mathbf{r}_{\mathrm{pd}},\mathbf{p})$ is the dipole coupling coefficient to mode $j$, which can be calculated using the Lorentz reciprocity theorem \cite{Lavrinenko2014}. The coupling coefficient depends on the dipole position $\mathbf{r}_{\mathrm{pd}}$ and dipole moment $\mathbf{p}$ through a dot product $\mathbf{p} \cdot \mathbf{E}_j(\mathbf{r}_{\mathrm{pd}})$. For the sake of notational clarity, we omit these dependencies in the following.  
Furthermore, $c_{j,\xi}$ are the expansion coefficients for mode $j$, and $\mathbf{g}_{\xi}(x_{\mathrm{p}},y_{\mathrm{p}})$ are the vectorial plane wave basis functions. 

The emitted field (\ref{eq:em_Ef}) consists of three contributions \cite{Snyder1983}: guided modes, radiating modes, and evanescent modes. In a waveguide surrounded by air, the eigenmode $j$ is guided if the propagation constant $\beta_j$ obeys $k_0^2 < \beta_j^2 \le (n_w k_0)^2$, where $n_w$ is the refractive index of the waveguide. In contrast the mode is radiating if $0<\beta_j^2 \le k_0^2$, and evanescent if $\beta_j^2 < 0$. We will apply this classification in Section \ref{sec:results} when we investigate the performance of the discretization schemes.
	
The normalized power emitted by a dipole to a selected mode can be expressed as \cite{Novotny2012Chap8}
\begin{eqnarray}
\nonumber \frac{P_j}{P_{\mathrm{Bulk}}} &=& \frac{\omega}{2} \frac{\mathrm{Im}\{ \textbf{p}^* \cdot a_j \mathbf{E}_j(\mathbf{r}_{\mathrm{pd}})  \}}{P_{\mathrm{Bulk}}} \\
 &=& \frac{\omega}{2} \frac{\mathrm{Im}\{ \textbf{p}^* \cdot \sum_{\xi} a_j c_{j,\xi} \mathbf{g}_{\xi}(\mathbf{r}_{\mathrm{pd}})\Delta k_{\xi}\}}{P_{\mathrm{Bulk}}}, 
\end{eqnarray}
where $P_{\mathrm{Bulk}} = |\textbf{p}|^2 n_B \omega^4 /(12\pi \epsilon_0 c^3)$ is the emitted power in a bulk medium of refractive index $n_B$. The normalized power is equal to the normalized spontaneous emission rate \cite{Novotny2012Chap8} $\gamma_j/\gamma_{\mathrm{Bulk}} = P_j/P_{\mathrm{Bulk}}$, where $\gamma_j$ and $\gamma_{\mathrm{Bulk}}$ are the spontaneous emission rates to the mode $j$ and to a bulk material, respectively. In the following we will only use the normalized unitless quantity $\Gamma_j = \gamma_j/\gamma_{\mathrm{Bulk}}$ for the emission rates.

\subsection{Reflection at an interface}

While the theory above holds for a structure with uniformity along the $z$ axis, most geometries of interest consist of several $z$-invariant sections. The full structure can be described by combining the solutions of Eqs. (\ref{eq:ecomp})-(\ref{eq:hcomp}) using a scattering matrix approach \cite{Li1996a,Lavrinenko2014}. Since our oFMM is based on expanding the fields in each layer using the same basis function, the reflections and transmission of the eigenmodes can be calculated conveniently using the expansion coefficients as described in the following.

Let $\mathbf{C}^{E}_{i}$ and $\mathbf{C}^{H}_{i}$, where $i=1,2$ is the layer index, be matrices whose columns contain the vector expansion coefficients for the lateral electric and magnetic fields respectively computed using (\ref{eq:ecomp}) and (\ref{eq:hcomp}). Then, at the interface of material layers 1 and 2, the transmission and reflection matrices are given as \cite{Lavrinenko2014}
\begin{eqnarray}
\mathbf{T}_{12} &=& 2\left[\left(\mathbf{C}^{E}_{1}\right)^{-1} \mathbf{C}^{E}_{2} + \left(\mathbf{C}^{H}_{1}\right)^{-1} \mathbf{C}^{H}_{2}\right]^{-1} \\ 
\mathbf{R}_{12} &=& \frac{1}{2}\left[\left(\mathbf{C}^{E}_{1}\right)^{-1} \mathbf{C}^{E}_{2} - \left(\mathbf{C}^{H}_{1}\right)^{-1} \mathbf{C}^{H}_{2}\right]\mathbf{T}_{12}.  
\end{eqnarray}

\section{Discretization scheme} \label{sec:disc}

We have now, via the modal representation in \eqref{eq:FourExpa2}, developed a formalism based on a non-uniform $k$-space discretization, which is a generalization of the uniform $k$-space discretization traditionally used in the Fourier modal method. In this section, we describe the important point of how to efficiently sample the $k$ space, before proceeding to example calculations

The lateral expansion basis function $g(k_x, k_y) = \exp[\ii (k_x x + k_y y) ]$ are plane waves defined entirely by the discretized values of the lateral wavenumbers $k_x$ and $k_y$. To discretize the transverse expansion basis efficiently in a general 3D approach, we generalize the non-uniform strategy used in the rotational symmetric case \cite{Hayrynen2016}. 

\begin{figure}[h]
\begin{center}
\includegraphics[width=0.65\columnwidth]
{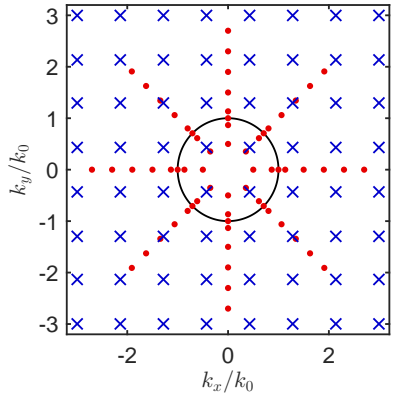}%
\caption{
Examples of the discrete mode distributions $\vekn{k}_\perp$ used with 3D oFMM. The blue crosses show the conventional equidistant discretization which may have different discretization step size for $x$ and $y$ directions, i.e. $\Delta k_x \neq \Delta k_y$. The red dots represent the dartboard discretization used with the open BC formalism. The solid line shows a unit circle $|\vekn{k}_\perp|/k_0=1$. In this simple example, we have used $k_{\mathrm{cut-off}}/k_0 = 3$ and 64 modes for both discretization schemes so that $N_x=N_y=8$, and $N_{\phi}=8$, $N_s=5$ and $\Delta k_{\mathrm{tail}}/k_0=0.4$.  
\label{Fig:KxKyDistributions_3DFMM}}
\end{center}
\end{figure}

First, in the conventional equidistant mode discretization approach, the spatial grid in $k$ space is given by
\begin{eqnarray}
(k_x^m, k_y^l) = (-k_{\mathrm{cut-off},x}+m\Delta k_x, -k_{\mathrm{cut-off},y}+l\Delta k_y), \label{eq:eqdis}
\end{eqnarray}
where $\Delta k_{\alpha} = 2k_{\mathrm{cut-off},{\alpha}}/(N_{\alpha}-1)$ and $m,l=0,\dots,(N_{\alpha}-1)$, with  $k_{\mathrm{cut-off},\alpha}$ being the cut-off value of the wavenumber and $N_{\alpha}$ the number of modes along the $\alpha=x,y$ axis, see Fig. \ref{Fig:KxKyDistributions_3DFMM}. In the following, when we apply the equidistant discretization scheme, we will use identical cut-off values $k_{\mathrm{cut-off},x}=k_{\mathrm{cut-off},y}$ and modes $N_x = N_y$ along the $k_x$ and $k_y$ axes.

Now, the proposed non-uniform circular non-separable discretization approach, which we in the following refer to as the "dartboard" scheme, is defined as follows. We consider the in-plane wavevector in polar coordinates and set $N_{\phi}$ rays on equidistantly placed angles, cf. Fig. \ref{Fig:KxKyDistributions_3DFMM}. Along each of the rays, the wavenumber values are sampled so that we use dense sampling in the interval $[0, 2k_0]$ symmetrically placed around $k_0$, and in the interval $[2k_0,k_{\mathrm{cut-off}}]$ a fixed step-size $\Delta k_{\mathrm{tail}}$ is used. The symmetric dense mode sampling is defined using a Chebyshev grid \cite{Boyd2001,DeLasson2012} as 
\begin{equation}
	\begin{array}{ll}
		k_m = k_0\sin(\theta_m),   &  1 \le m \le N_s/2 \\ 
		k_m = k_0[2-\sin(\theta_m)],  & N_s/2+1 \le m \le N_s,
	\end{array}
\end{equation}
where $\theta_m = \frac{m \pi}{N_s+1}$ and $N_s$ is the number of modes in the interval $[0, 2k_0]$.
Thus, in the dartboard discretization approach we have four parameters $N_{\phi}$, $N_s$, $\Delta k_{\mathrm{tail}}$, and $k_{\mathrm{cut-off}}$.
The motivation of using symmetric dense sampling around $k_0$ is to accurately account for the radiating modes as discussed in detail in \cite{Hayrynen2016}. In the next section, we show that the dartboard discretization approach outperforms the conventional equidistant mode sampling. As pointed out in \cite{Hayrynen2016}, the dartboard mode sampling approach described here is not necessarily the universally optimal, and geometry specific variations may be adopted instead. However, with the proposed approach significant improvement is achieved in terms of the required number of modes and thus of the required computational power.   

\section{Results} \label{sec:results}

Next, after introducing the principles of the oFMM formalism and the efficient mode sampling scheme, we test our method by investigating its performance for the two cases of light emission by a dipole in a square waveguide as well as of reflection at a waveguide-metal interface. Both examples depend critically on a correct and accurate description of the open BC. We also compare the new discretization scheme to the conventional discretization used in connection with Li's factorization rules. As already mentioned, in Appendix \ref{app:deriv} we show how Li's factorization rules are correctly used with oFMM based on equidistant discretization, whereas the equations implementing the non-separable dartboard discretization used in this manuscript are not compatible with the inverse factorization rule. However, our results will demonstrate that, even without the inverse factorization rule the dartboard discretization approach outperforms the equidistant discretization implemented using Li's factorization rules.

\subsection{Dipole emission in a square waveguide}

We first investigate light emission in a square waveguide by calculating the emission rates to the guided modes and to the radiation modes. Additionally, we compute the spontaneous emission factor $\beta$ (not to be confused with the propagation constant $\beta_j$) describing the ratio of emitted light coupled to the fundamental guided mode. While typical nanophotonic waveguides support only a few guided modes, the total emission rate and thus the $\beta$ factor depend on the emission into the continuum of radiation modes leaking out of the waveguide. The strength of the oFMM method becomes apparent when determining the light emission to the radiation modes. 

\begin{figure}[h!] 
\centering
\includegraphics[width=0.9\columnwidth]
{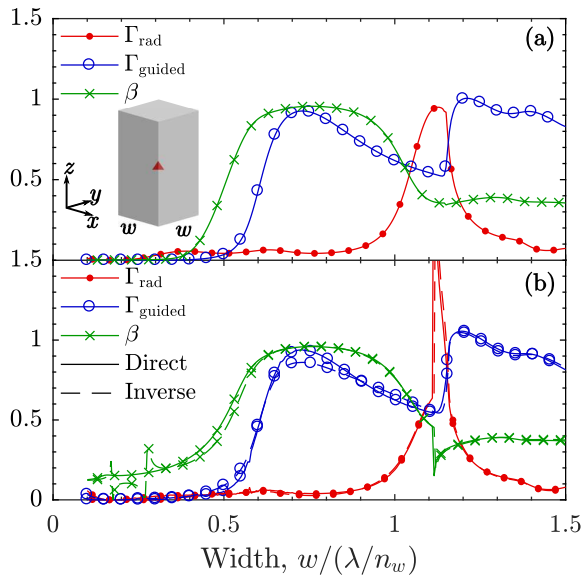}
\caption{Emission from a point dipole placed on-axis of an infinitely long square waveguide having widths $w_x=w_y=w$. The dipole is oriented along the $x$ axis. 
(a) The normalized emission to the radiation modes, to the guided modes, and the $\beta$ factor calculated using the dartboard discretization scheme with $N_{\phi} = 14$, $N_s=180$, $k_{\mathrm{cut-off}}/k_0=15$, and $\Delta k_{\mathrm{tail}}/k_0=0.06$.
(b) The corresponding data calculated using conventional square sampling and applying both the direct and Li's inverse factorization rules with $N_x=N_y = 80$ and $k_{\mathrm{cut-off}}/k_0=15$. 
The wavelength used in the calculations is $\lambda$ = 1 $\mu$m. The total number of modes are (a) 5558, and (b) 6400. 
}\label{Fig:WGem}
\end{figure}

Similar to the investigations presented in \cite{Lecamp2006}, we consider a dipole emitter oriented along the $x$ axis placed on the axis of an infinitely long square waveguide with varying edge length $w_x=w_y$ and refractive index $n_w = 3.5$ surrounded by air. Figure \ref{Fig:WGem}(a) presents the $\beta$ factor and the emission rates to the guided modes and to the radiation modes as functions of the waveguide size calculated using the dartboard discretization. The rates are normalized to the bulk emission rate (see Section \ref{sec:theory}\ref{sec:emission}). 
Figure \ref{Fig:WGem}(b) shows the same properties of the waveguide calculated using an equidistant square sampling using either the direct or the inverse factorization rules. 
The emission rate to the guided modes calculated with the three approaches agree well, and in contrast, a clear difference is seen in the emission rate to the radiation modes and therefore also in the $\beta$ factor. In particular, the coupling to the radiation modes exhibit a spike around a normalized width of 1.15 with the square sampling, which gives an unphysical kink in the $\beta$ factor.
The discretization parameters used in Fig. \ref{Fig:WGem} are given in the figure caption and were selected as a result of the convergence investigations presented in the following.  

\begin{figure}[h!] 
	\centering
	\includegraphics[width=0.9\columnwidth]
	{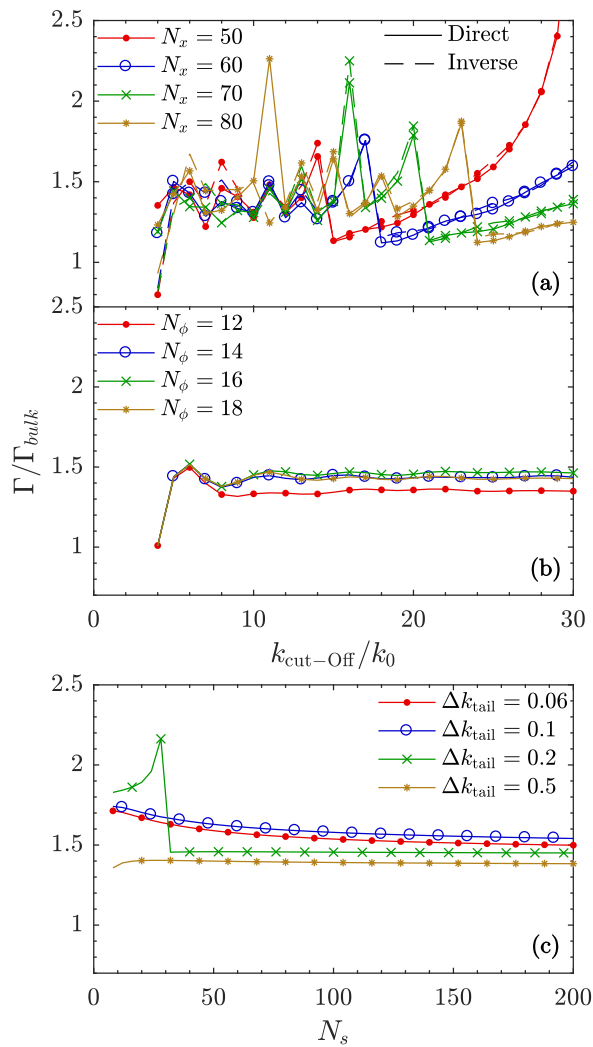}
	\caption{Convergence comparison of the total emission rate using the three approaches for a waveguide $w/(\lambda/n_{w}) = 1.15$.
		(a) The total emission rate for the equidistant discretization as a function of the cut-off value computed with the number of modes shown in the legend using the direct factorization rule (solid line) and the inverse factorization rule (dashed line).   
		(b) The emission rate as a function of the cut-off value computed using the dartboard mode sampling and the number of angles shown in the legend with $N_s=140$ modes on the symmetric radial part with $\Delta k_{\mathrm{tail}}/k_0=0.2$. 
		(c) The emission rate as function of the number of modes $N_s$ in the symmetric sampling part computed using the dartboard mode sampling with a fixed number of angles $N_{\phi}=16$ and cut-off value $k_{\mathrm{cut-off}}/k_0=13$. 
	}
	\label{Fig:WGconv}
\end{figure}

To further investigate the performances of the three approaches we fix the waveguide geometry by setting the width to $w=1.15\lambda/n_w$ and vary the cut-off value of the transverse wavenumber as well as the number of modes. This waveguide size is selected for the convergence investigations since a clear difference of the results is seen for this diameter in Fig. \ref{Fig:WGem}. Figures \ref{Fig:WGconv}(a,b) show the convergence investigations of the total emission rate as a function of the cut-off value with several different mode numbers, while Fig. \ref{Fig:WGconv}(c) shows the convergence of the total emission rate as a function of the number of modes $N_s$ in the interval $[0,2k_0]$ for the dartboard discretization scheme. The dartboard approach shows clear convergence around a cut-off of $\approx 15k_0$. In contrast, the equidistant discretization scheme does not guarantee convergence even with cut-off value of $30k_0$. 
The maximum number of modes used in the calculations of Fig. \ref{Fig:WGconv}(a) are on the upper limit of the performance of our HPC cluster computer. This is also the case for the highest number of modes used in Figs. \ref{Fig:WGconv}(b,c). However, in Figs. \ref{Fig:WGconv}(b,c) the convergence is achieved also for the cases with smaller number of modes.       

When using the equidistant discretization, numerical artifacts in the form of large oscillations are observed at particular values of the number of modes and cut-off as displayed in Figs. \ref{Fig:WGem}(b) and \ref{Fig:WGconv}(a) (as well as in Figs. \ref{Fig:WG_MM}(b) and \ref{Fig:WG_MM_conv}(a)). 
As discussed in Appendix \ref{app:pbc}, the oFMM together with the equidistant square discretization scheme mathematically corresponds to having periodic BCs and to using a Fourier series expansion, where the periodic lengths of the computational domain are inversely proportional to $\Delta k_x$ and $\Delta k_y$. For geometries with periodic BCs, destructive or constructive interference due to light emission in the neighboring periodic elements may occur leading to the observed large oscillations of the emission rates, that thus are an inherent consequence of the equidistant discretization scheme. 

A common approach to circumvent these artifacts due to periodic BCs is to use artificial absorbing BCs, often in the form of the so-called perfectly matched layers (PMLs) \cite{Berenger1994,Hugonin2005a}. However, for the modal method with a PML BC, the convergence of the emission properties with the PML parameters towards the open geometry limit \cite{Gregersen2008a, Gregersen2010} is not well-established with errors in some cases as high as $\approx$ 20 \% \cite{DeLasson2015a}.
In contrast, the oFMM with the efficient discretization scheme relies on a truly open computational domain, and therefore avoids using artificial or periodic BCs leading to improved accuracy and convergence towards the true open geometry limit.      

\subsection{Reflection from dielectric waveguide-metal interface}

As a second example, we investigate convergence of the method for a structure consisting of an infinite waveguide standing on top of a metallic mirror by computing the reflection coefficient of the fundamental guided mode from the waveguide-metal interface. The refractive indices of the waveguide and metal are $n_w=3.5$ and $n_{\mathrm{Ag}}= \sqrt{-41 + 2.5\iim}$ at the wavelength $\lambda$ = 1 $\mu$m. 

\begin{figure}[h!] 
\centering
\includegraphics[width=0.9\columnwidth]
{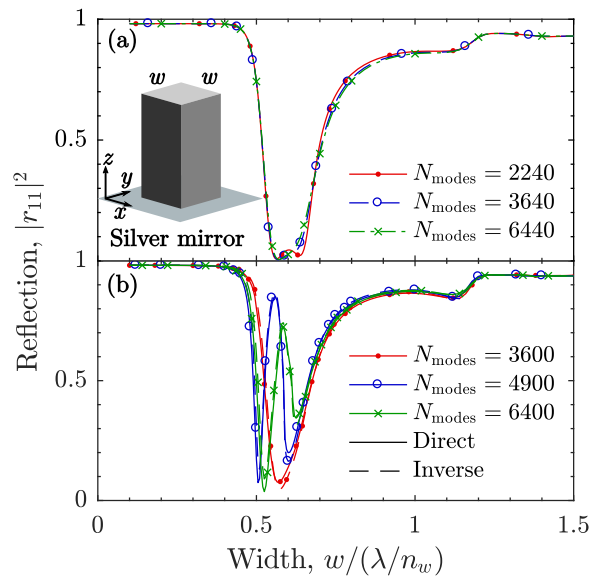}
\caption{The reflection of the fundamental waveguide mode from a metal mirror calculated using 
(a) the darboard discretization and (b) the equidistant discretization.
The cut-off in both cases is $k_{\mathrm{cut-off}}/k_0 = 14$, and for the dartboard discretization fixed values of $N_{\phi} = 14$ and $\Delta k_{\mathrm{tail}}/k_0 = 0.2$ were used.
The legends show the total number of modes used.}
\label{Fig:WG_MM}
\end{figure}

Figure \ref{Fig:WG_MM} shows the calculated reflection coefficient as a function of the waveguide size $w_x=w_y$ using (a) the dartboard sampling and (b) the equidistant discretization employing the direct and inverse factorization rules with several different number of discretization modes. 
The cut-off in all cases is $k_{\mathrm{cut-off}}/k_0 = 14$. Furthermore, for the dartboard discretization fixed values of $N_{\phi} = 14$ and $\Delta k_{\mathrm{tail}}/k_0 = 0.2$ were used and only $N_s$ was varied. These parameters were chosen to achieve convergence according to the investigations discussed in the next paragraph.
In narrow waveguides, the reflection coefficients are essentially determined by the air-metal reflection  ($R_{\mathrm{Air-Ag}} \approx 0.98$) since in this limit the fundamental mode is mainly localized in the air surrounding the waveguide. In contrast, in the limit of large waveguides the fundamental mode is primarily confined in the GaAs waveguide ($R_{\mathrm{GaAs-Ag}} \approx 0.95$). 
A dramatic difference in the results is seen in the region around $w/(\lambda/n_w) \approx 0.6$, where the reflectivity drops due to a surface-plasmon mediated coupling predominantly to radiation modes propagating in directions perpendicular to the waveguide axis \cite{Friedler2008}. When a substantial amount of light is propagating in the $x$-$y$ plane, the performance of the open boundary condition becomes critical, and comparison of Figs. \ref{Fig:WG_MM}(a) and \ref{Fig:WG_MM}(b) clearly demonstrates that this light emission is better resolved using the dartboard discretization.

\begin{figure}[h!] 
\centering
\includegraphics[width=0.9\columnwidth]
{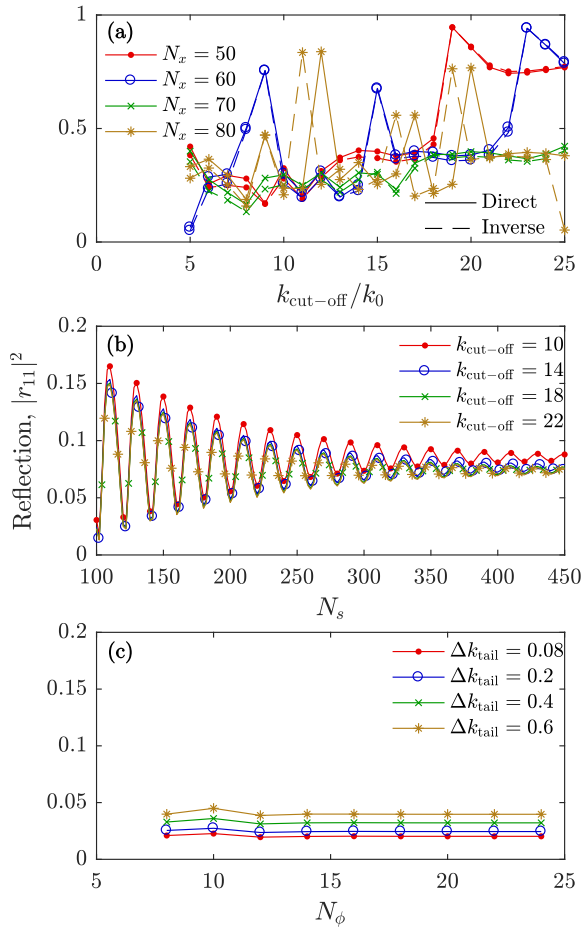}
\caption{Convergence of the reflection of the fundamental waveguide mode from a metal mirror.
(a) The reflection as function of the cut-off value with number of modes shown in the legend using the direct factorization rule (solid line) and the inverse factorization rule (dashed line).  
(b) The reflection as function of the number of modes in the symmetric sampling part using the dartboard mode sampling with fixed $N_{\phi}=14$ and $\Delta k_{\mathrm{tail}}/k_0 = 0.2$, and cut-off value shown in the legend. 
(c) The reflection as function of the number of angles using the dartboard mode sampling with fixed $k_{\mathrm{cut-off}}/k_0=14$ and $N_s=100$, and $\Delta k_{\mathrm{tail}}/k_0$ shown in the legend. Note the different scaling between (a) and (b-c).
}
\label{Fig:WG_MM_conv}
\end{figure}

Whereas the reflection coefficients in Figs. \ref{Fig:WG_MM}(a) and (b) are obtained for a fixed cut-off value, we now fix the geometry and study the effect of the cut-off value of $k_m$. We select a waveguide width of $w_x=w_y=0.63\lambda/n_w$, since Fig. \ref{Fig:WG_MM}(b) reveals this to be a challenging computational point. The convergence investigation is shown in Fig. \ref{Fig:WG_MM_conv}. The dartboard discretization (Figs. \ref{Fig:WG_MM_conv}(b,c) again leads to convergence with respect to all of the four discretization parameters. In contrast, no clear convergence is seen when using the equidistant discretization, while we also in this case approach the performance limit of our HPC cluster computer. As discussed in the previous section, the peaks observed in Figs. \ref{Fig:WG_MM}(b) and \ref{Fig:WG_MM_conv}(a) are a consequence of the periodicity of the computational domain when using the equidistant discretization scheme.

\section{Discussion}\label{sec:discussion}

The convergence checks in the selected waveguide examples presented in Figs. \ref{Fig:WGconv} and \ref{Fig:WG_MM_conv} show that our method converges for the investigated waveguide sizes and structures. The non-separable nature of our discretization scheme prevents the use of Li's factorization rules, but even when using the standard direct factorization, a clear improvement in the performance is obtained using the proposed dartboard discretization scheme compared to the conventional equidistant discretization of the basis functions. Although these examples do not guarantee the convergence of our method for all imaginable waveguide sizes and geometries, we generally expect our method to deliver improved performance for various types of waveguides, possibly with additional geometry specific modifications to the discretization scheme. 

In high-index-contrast structures as the examples presented here, the FMM method, due to the difficulty of resolving large discontinuities using a plane wave expansion, generally requires a significant amount of modes to achieve convergence. Whereas this may not be a computational difficulty in a rotational symmetric case which in the lateral plane reduces to a 1D problem, the size of the eigenvalue problem in the general planar 2D case rapidly explodes when the number of modes are increased \cite{DeLasson2015a}. Thus, we expect that a further improvement in terms of computational efficiency could be obtained by combining the dartboard discretization scheme with an adaptive spatial coordinate scheme \cite{Essig2010} or by introducing a semi-analytical approach for defining the eigenmodes. In the rotationally symmetric case, exact analytical descriptions of the eigenmodes exist \cite{Snyder1983}, while in the rectangular case approximate solutions \cite{Raghuwanshi2009} could be used. 

\section{Conclusion}\label{sec:conc}

We have generalized the recently reported open geometry Fourier modal method formalism relying on open boundary conditions and a non-uniform circular "dartboard" $k$-space sampling for general 3D systems, allowing e.g. the modeling of rectangular waveguides. By applying open boundary conditions, we avoid using the artificial absorbing BCs. We have demonstrated the efficiency of the approach by investigating dipole emission in a square waveguide structure and by studying the reflection coefficient of the fundamental waveguide mode for a waveguide-metal mirror interface, that both are problems of fundamental interest when designing nanophotonic devices.
We expect that our new method will prove useful in accurate modeling of a variety of nanophotonic structures, for which correct treatment of an open boundary is crucial. 

\section*{Acknowledgment}
Support from the Danish Research Council for Technology and Production via the Sapere Aude project LOQIT (DFF - 4005-00370) and support from the Villum Foundation via the VKR Centre of Excellence NATEC are gratefully acknowledged.

\appendix

\section{Derivation of the eigenvalue problem in 3D open geometry} \label{app:deriv}

\subsection{Fourier expansion of the field components}

The vector components of Maxwell's equations in Cartesian coordinates are \cite{Lavrinenko2014}:
\begin{align}
\partial_y E_z \mp \ii \beta E_y &= \ii \omega \mu_0 H_x, \label{eqn:MaxHx} \\
\pm \ii \beta E_x - \partial_x E_z &= \ii \omega \mu_0 H_y, \label{eqn:MaxHy} \\
\partial_x E_y - \partial_y E_x &= \ii \omega \mu_0 H_z, \label{eqn:MaxHz} \\
\partial_y H_z \mp \ii \beta H_y &= -\ii \omega \varepsilon_0 \varepsilon E_x, \label{eqn:MaxEx} \\
\pm \ii \beta H_x - \partial_x H_z &= -\ii \omega \varepsilon_0 \varepsilon E_y, \label{eqn:MaxEy} \\
\partial_x H_y - \partial_y H_x &= -\ii \omega \varepsilon_0 \varepsilon E_z, \label{eqn:MaxEz}
\end{align}
where the harmonic time-dependence $\exp(-\ii \omega t)$ of the fields is assumed and the propagation along the $z$ axis is treated analytically as $\exp (\pm \ii \beta z)$.
Thus, the field components in a uniform layer only depend on the lateral coordinates $(x,y)$ and are represented as
\begin{equation} \label{eqn:FunctionExpansion}
f(x,y) = \int_{-\infty}^{\infty} \int_{-\infty}^{\infty} c_f(k_x, k_y) g(k_x, k_y) \dd k_x \dd k_y.
\end{equation}
The basis function $g(k_x, k_y) = \exp[\ii (k_x x + k_y y) ]$ are plane waves and satisfy the following orthogonality condition
\begin{equation} \label{eqn:Orthogonality}
\int_{-\infty}^{\infty} \int_{-\infty}^{\infty} g(k_x, k_y) g^*(k_x', k_y') \dd x \dd y = (2\pi)^2 \delta(k_x - k_x') \delta(k_y - k_y').
\end{equation}
The expansion coefficients in Eq. (\ref{eqn:FunctionExpansion}) are obtained by multiplying with $g^*(k_x', k_y')$, integrating over the transverse plane and using the orthogonality relation (\ref{eqn:Orthogonality}) leading to 
\begin{equation} \label{eqn:cf}
c_f (k_x, k_y) = \frac{1}{(2\pi)^2} \int_{-\infty} ^{\infty} \int_{-\infty} ^{\infty} f (x,y) g^*(k_x, k_y) \dd x \dd y.
\end{equation}

The material properties of the structure are described by the permittivity and impermittivity functions, which are written as a sum between a constant background value and a position dependent deviation from the background value, as
\begin{align}
\varepsilon(x,y) &= \varepsilon_B + \Delta\varepsilon(x,y), \\
\eta(x,y) &= \frac{1}{\varepsilon(x,y)} = \eta_B + \Delta \eta(x,y),
\end{align}
where $\Delta \varepsilon(x,y)$ and $\Delta \eta(x,y)$ are functions with compact support, such that $\Delta\varepsilon = \Delta \eta = 0$ outside a finite domain. 

The expansion coefficients of the Fourier transform of the permittivity function can then be written as
\begin{eqnarray}
c_{\varepsilon}(k_x, k_y) = \varepsilon_B \delta(k_x) \delta(k_y) + c_{\Delta\varepsilon}(k_x, k_y) \label{eq:epsex1}
\end{eqnarray}
where
\begin{eqnarray}
c_{\Delta\varepsilon}(k_x, k_y) =  \frac{1}{(2\pi)^2} \int_{-\infty} ^{\infty} \int_{-\infty} ^{\infty} \Delta\varepsilon(x,y) g^*(k_x, k_y) \dd x \dd y. \label{eq:epsex2}
\end{eqnarray}
The Fourier transforms of the position dependent deviations, $\Delta \varepsilon$ and $\Delta \eta$, are thus obtained by calculating finite integrals, whereas the constant $\varepsilon_B$ and $\eta_B$ contributions are handled analytically using Dirac delta functions.

In order to factorize Eqs. (\ref{eqn:MaxHx})-(\ref{eqn:MaxEz}) by insertion of the expansion in Eq. (\ref{eqn:FunctionExpansion}), Li's factorization rules \cite{Li1997, Li1996b, Lalanne1998} should be considered. Eqs. (\ref{eqn:MaxHx})-(\ref{eqn:MaxHz}) and Eq. (\ref{eqn:MaxEz}) do not contain any products between two functions with concurrent jumps (discontinuities) and therefore the direct rule applies in these equations. However, in Eq. (\ref{eqn:MaxEx}) and (\ref{eqn:MaxEy}) the product $\varepsilon E_{x,y}$ is continuous, but both $\varepsilon(x,y)$ and $E_{x,y}(x,y)$ are discontinuous functions, thus they have concurrent jumps, and the inverse factorization rule should - ideally - be used.

\subsection{Direct factorization rule}

We start by factorizing 
Eqs. (\ref{eqn:MaxHx})-(\ref{eqn:MaxEz}) and writing them in a matrix form one-by-one. Inserting the function expansion in Eq. (\ref{eqn:FunctionExpansion}) into Eq. (\ref{eqn:MaxHx}) leads to
\begin{align}
\int \int \left[ k_y c_{E_z}(k_x, k_y) \mp \beta c_{E_y}(k_x, k_y) \right] g(k_x, k_y) \dd k_x \dd k_y  = \nonumber \\
\omega \mu_0 \int \int c_{H_x}(k_x, k_y) g(k_x, k_y) \dd k_x \dd k_y,
\label{eq:drf}
\end{align}
where the integration limits (from $-\infty$ to $\infty$) have been omitted for notational clarity. Multiplying with $g^*(k_x', k_y')$, integrating over $x$ and $y$ and using the orthogonality relation (\ref{eqn:Orthogonality}) lead to
\begin{eqnarray}
\nonumber
&& (2\pi)^2 \int \int \left[ k_y c_{E_z}(k_x, k_y) \mp \beta c_{E_y}(k_x, k_y) \right] \\
\nonumber
&& \quad \times \delta(k_x - k_x') \delta(k_y - k_y') \dd k_x \dd k_y \\ 
\nonumber
&=& (2\pi)^2 \omega \mu_0 \int \int c_{H_x}(k_x, k_y) \\
&& \quad \times \delta(k_x - k_x') \delta(k_y - k_y')\dd k_x \dd k_y.
\label{eq:dfi}
\end{eqnarray}
Performing the integrations in Eq. (\ref{eq:dfi}) we arrive at
\begin{equation}
k_y c_{E_z}(k_x, k_y) \mp \beta c_{E_y}(k_x, k_y) = \omega \mu_0 c_{H_x}(k_x, k_y),
\label{eq:drff}
\end{equation}
which after discretization of the $k$ space is written in matrix form as
\begin{equation} \label{eqn:EyMatrix}
\mathbf{k}_y \mathbf{e}_z \mp \beta \mathbf{e}_y = \omega \mu_0 \mathbf{h}_x,
\end{equation}
where $\mathbf{e}_y$ is a vector with $c_{E_y}^\xi$ as elements. $\mathbf{k}_x$ and $\mathbf{k}_y$ are diagonal matrices with elements $k_x^\xi$ and $k_y^\xi$.

Using a similar approach, Eqs. (\ref{eqn:MaxHy}) and (\ref{eqn:MaxHz}) are written in matrix form as
\begin{align} \label{eqn:ExMatrix}
\pm \beta \mathbf{e}_x - \mathbf{k}_x \mathbf{e}_z &= \omega \mu_0 \mathbf{h}_y, \\
\mathbf{k}_x \mathbf{e}_y - \mathbf{k}_y \mathbf{e}_x &= \omega \mu_0 \mathbf{h}_z. \label{eqn:HzMatrix}
\end{align}

Next we prepare Eq. (\ref{eqn:MaxEz}) in a discretized form in order to eliminate $\mathbf{e}_z$ from Eqs. (\ref{eqn:EyMatrix}) and (\ref{eqn:ExMatrix}), which can also be performed by applying the direct factorization rule. Expanding the field components, using (\ref{eq:epsex1})-(\ref{eq:epsex2}) and performing a change of variables $\hat{k}_{x,y} = k_{x,y} + k_{x,y}'$ lead to
\begin{eqnarray}
\nonumber
&& \int \int \left[ k_x c_{H_y}(k_x, k_y) - k_y c_{H_x}(k_x, k_y) \right]  \\
\nonumber
&& \quad\times \exp \left[ \ii \left(k_x x + k_y y \right) \right] \dd k_x \dd k_y \\
\nonumber
&=& - \omega \varepsilon_0 \int \int \int \int \bigg[ \varepsilon_B \delta(\hat{k}_x - k_x) \delta(\hat{k}_y - k_y)\\
\nonumber
&& \qquad + c_{\Delta\varepsilon}(\hat{k}_x - k_x, \hat{k}_y - k_y) \bigg]c_{E_z}(k_x, k_y) \\
&& \quad \times  \exp \left[ \ii \left(\hat{k}_x x + \hat{k}_y y \right) \right] \dd k_x \dd k_y \dd \hat{k}_x \dd \hat{k}_y.
\label{eq:ToepExp}
\end{eqnarray}
We then multiply with $\exp[-\ii(k_x' x + k_y' y)]$, integrate over $x$ and $y$ and employ the orthogonality condition (\ref{eqn:Orthogonality}) and obtain
\begin{eqnarray}
\nonumber
&& k_x' c_{H_y}(k_x', k_y') - k_y' c_{H_x}(k_x', k_y') \\
\nonumber
&=& - \omega \varepsilon_0 \int \int \bigg[ \varepsilon_B \delta(k_x' - k_x) \delta(k_y' - k_y) \\
&& \quad + c_{\Delta\varepsilon}(k_x' - k_x, k_y' - k_y) \bigg] c_{E_z}(k_x, k_y) \dd k_x \dd k_y.
\label{eq:ezd} 
\end{eqnarray}
In discretized form Eq. (\ref{eq:ezd}) is written as
\begin{equation}
\mathbf{k}_x \mathbf{h}_y - \mathbf{k}_y \mathbf{h}_x = -\omega \varepsilon_0 \left[ \bm{\Delta\varepsilon} \Delta \mathbf{k} + \varepsilon_B \mathbf{I} \right] \mathbf{e}_z, \label{eq:epsmdef}
\end{equation}
where $\bm{\Delta\varepsilon}$ is the Toeplitz matrix containing the elements  $c_{\Delta\varepsilon}^\xi = c_{\Delta\varepsilon}(k_x^\xi, k_y^\xi)$, $\mathbf{I}$ is the identity matrix and $\Delta \mathbf{k}$ is the diagonal matrix containing the discretized area elements $\Delta k_\xi$ in $k$ space. Thus, $\mathbf{e}_z$ equals to
\begin{equation}
\mathbf{e}_z = -\frac{1}{\omega \varepsilon_0} \left[ \bm{\Delta\varepsilon} \Delta \mathbf{k}  + \varepsilon_B \mathbf{I} \right]^{-1} \left[  \mathbf{k}_x \mathbf{h}_y - \mathbf{k}_y \mathbf{h}_x \right].
\end{equation}
allowing us to write Eqs. (\ref{eqn:EyMatrix}) and (\ref{eqn:ExMatrix}) in the form of an eigenvalue problem that couples the lateral electric field components to the lateral magnetic field components as 
\begin{equation}
\left[ \begin{matrix}
\mathbf{k}_x \bm{\varepsilon}_{\mathrm{Tot}}^{-1} \mathbf{k}_y & -\mathbf{k}_x \bm{\varepsilon}_{\mathrm{Tot}}^{-1} \mathbf{k}_x + k_0^2 \mathbf{I} \\
\mathbf{k}_y \bm{\varepsilon}_{\mathrm{Tot}}^{-1} \mathbf{k}_y - k_0^2 \mathbf{I} & -\mathbf{k}_y \bm{\varepsilon}_{\mathrm{Tot}}^{-1} \mathbf{k}_x
\end{matrix} \right] \left[ \begin{matrix}
\mathbf{h}_x\\
\mathbf{h}_y
\end{matrix} \right] = \pm \omega \varepsilon_0 \beta \left[ \begin{matrix}
\mathbf{e}_x \\
\mathbf{e}_y
\end{matrix}\right]
\label{eq:ExEyMat}
\end{equation}
where $ \bm{\varepsilon}_{\mathrm{Tot}} =  \bm{\Delta\varepsilon} \Delta \mathbf{k} + \varepsilon_B \mathbf{I}$.

From Eqs. (\ref{eqn:MaxEx}) and (\ref{eqn:MaxEy}) we can write similar set of equations that couples the lateral components so that Eqs. (\ref{eqn:MaxEx}) and (\ref{eqn:MaxEy}) together with Eq. (\ref{eq:ExEyMat}) allows us to eliminate the magnetic field components and form an eigenvalue problem for the lateral electric field components (or vice versa). 
However, Eqs. (\ref{eqn:MaxEx}) and (\ref{eqn:MaxEy}) need special treatment due to the product $\varepsilon E_{x,y}$. 

\subsection{Inverse factorization approach}

In the following the application of the inverse rule for open boundaries with a separable discretization grid in $k$ space will be presented. As discussed in Appendix \ref{app:pbc}, an equidistant discretization with an open BC is mathematically equivalent to implementing a periodic BC and a Fourier series expansion. Furthermore, as will become apparent in the course of deriving the inverse factorization for the separable discretization, the inverse factorization approach is not applicable for our dartboard discretization scheme defined in Section \ref{sec:disc}.   

\begin{figure}[h!]
\centering
\includegraphics[width=0.5\columnwidth]
{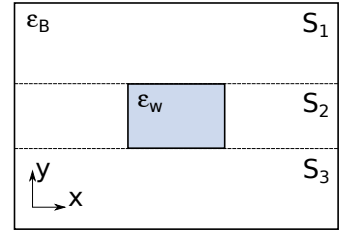}
\caption{A waveguide in air is divided into three sections separated by the $y$ interfaces of the permittivity function. Here the background permittivity is $\varepsilon_B$ and in the waveguide region $\Delta\varepsilon(x,y) = \varepsilon_w-\varepsilon_B$. The permittivity is $y$ independent inside each of the three sections.}
\label{fig:WaveguideInAir}
\end{figure}

The factorization will be performed on Eq. (\ref{eqn:MaxEx}) to illustrate how the inverse rule is implemented for the product $\varepsilon E_x$. 
The matrix representation for the $\varepsilon$ function used in the product $\varepsilon E_x$ will be denoted $\bm{\varepsilon}_x$, indicating that it accommodates for continuity of the product along the $x$ direction, where the inverse rule is applied as in \cite{Li1997, Lalanne1998}. 
Now, $E_x$ is discontinuous in the $x$ direction but continuous in the $y$ direction. $\varepsilon$ is discontinuous in both the $x$ and $y$ direction. Their product, $\varepsilon E_x$, is continuous in the $x$ direction and discontinuous in the $y$ direction, thus the inverse rule is used for the $x$ direction and the direct rule for the $y$ direction. The way this is done computationally is to divide the structure into sections separated by the interfaces in the $y$ direction and apply the inverse rule to each of these sections. This is illustrated in Figure \ref{fig:WaveguideInAir}.

In general the expansion coefficients for all $(x, y)$-dependent functions are given as in Eq. (\ref{eqn:cf}). The integration over the $y$ coordinate is then separated into sections where the function is uniform along the $y$ axis. Using Figure \ref{fig:WaveguideInAir} as the example, the $y$ integration is separated into three parts
\begin{align}
c_f (k_x, k_y) = &\frac{1}{2\pi} \int_{S_1} f_{x,S_1}(k_x) \exp(- \ii k_y y) \dd y \nonumber \\
+ &\frac{1}{2\pi} \int_{S_2} f_{x,S_2}(k_x) \exp(- \ii k_y y) \dd y \nonumber \\
+ &\frac{1}{2\pi} \int_{S_3} f_{x,S_3}(k_x) \exp(- \ii k_y y) \dd y , \label{eq:ifdiv}
\end{align}
where 
\begin{equation}
f_{x,S_i}(k_x) = \frac{1}{2\pi} \int f(x,y_{S_i}) \exp(- \ii k_x x) \dd x.
\end{equation}
Here the notation $f(x,y_{S_i})$ means that the function is evaluated within section $S_i$ and is only dependent on the $x$ coordinate within that section. With this separation, it is possible to factorize $\varepsilon$ using the correct factorization rules \textit{provided that the discretized basis set features separable $k_x$ and $k_y$ dependency} as in \eqref{eq:eqdis}. If this is the case, we can index the $k_x$ and $k_y$ contributions to the basis mode $k$ vector as $(k_x^m,k_y^l)$ using separate indices $m$ and $l$. It is then possible to apply the inverse rule to the product $\varepsilon_x E_x$ factorized along the $x$ direction by first preparing the Fourier transform along the $x$ axis of the inverse permittivity as
\begin{eqnarray}
\nonumber
\eta_{x,S_i} (k_x)   %_{m-n} 
&=& \frac{1}{2\pi} \int_{-\infty}^{\infty} \Delta \eta(x, y_{S_i}) \exp(-\ii k_x x) \dd x  \\
&&   + \eta_B \delta(k_x). \label{eq:etaexp}
\end{eqnarray}
We then form the Toeplitz matrix for the $\eta_{x,S_i}$ function discretized on the $k_x^m$ grid. Since the product of the expansions of $\varepsilon$ and $E_x$ involves an integration over $k$ space as in \eqref{eq:ezd}, the Toeplitz matrix is given by
\begin{eqnarray}
	\bm{\eta}_{x,S_i,\mathrm{Tot}} = \bm{\Delta\eta}_{x,S_i}\Delta \mathbf{k} _x + \eta_B \bm{I}, \label{eq:etadisc}
\end{eqnarray}
where $\bm{\Delta\eta}_{x,S_i}$ is the Toeplitz matrix containing the elements $\Delta\eta_{x,S_i}^m=\Delta\eta_{x,S_i}(k_x^m)$ and $\Delta \mathbf{k}_x$ is the diagonal matrix with $\Delta k_x^m$ as elements. According to the inverse rule, we then take the inverse of this matrix and Fourier transform the resulting elements along the $y$ axis as
\begin{eqnarray} %\label{eqn:EpsX}
\nonumber \varepsilon_{x,mn} (k_y) & = &\frac{1}{2\pi} \int_{-\infty}^{\infty} (\bm{\Delta\eta}_{x,\mathrm{Tot}}^{\mathrm{Inv}})_{mn}(y) \exp \left(- \ii k_y y \right) \dd y \\ 
&& + \varepsilon_B \delta_{mn} \delta(k_y),
\end{eqnarray}
where $\bm{\Delta\eta}_{x,\mathrm{Tot}}^{\mathrm{Inv}} (y) =  \bm{\eta}_{x,S_i,\mathrm{Tot}}^{-1} - \varepsilon_B\bm{I}$, which is piece-wise constant over the various regions $S_i$ as discussed above. 

The final Toeplitz matrix $\bm{\varepsilon}_x$ is then obtained by introducing the discretization on the $k_y^l$ grid, and its elements are given by 
\begin{eqnarray} \label{eqn:EpsX}
\nonumber (\bm{\varepsilon}_x)_{mn, lj} &=& \frac{1}{2\pi} \int_{-\infty}^{\infty} (\bm{\Delta\eta}_{x,\mathrm{Tot}}^{\mathrm{Inv}})_{mn}(y) \exp \left(- \ii (k_y^l - k_y^j) y \right) \dd y \Delta k_y^j \\
&& + \delta_{mn}\delta_{lj} \varepsilon_B.
\end{eqnarray}
Similarly for the product $\varepsilon_y E_y$ we obtain
\begin{eqnarray} \label{eqn:EpsY}
\nonumber (\bm{\varepsilon}_y)_{mn, lj} &=& \frac{1}{2\pi} \int_{-\infty}^{\infty} (\Delta\bm{\eta}_{y,\mathrm{Tot}}^{\mathrm{Inv}})_{lj}(x) \exp \left( -\ii (k_x^m - k_x^n) x \right) \dd x  \Delta k_x^n \\
&& + \delta_{mn}\delta_{lj} \varepsilon_B.
\end{eqnarray}
The integrals in Eqs. (\ref{eqn:EpsX}) and (\ref{eqn:EpsY}) can be carried out analytically when the matrix $\Delta\bm{\eta}_{x(y),\mathrm{Tot}}^{\mathrm{Inv}}$ has been found for each $y(x)$-independent section.

The factorization of Eqs. (\ref{eqn:MaxEx})--(\ref{eqn:MaxEy}) thus become
\begin{eqnarray}
\ii k_y \mathbf{h}_z \mp \ii \beta \mathbf{h}_y &=& -\ii\omega \varepsilon_0 \bm{\varepsilon}_x \mathbf{e}_x \\
\mp \ii \beta \mathbf{h}_x - \ii k_x \mathbf{h}_z  &=& -\ii\omega \varepsilon_0 \bm{\varepsilon}_y \mathbf{e}_y.
\end{eqnarray}
Eliminating $\mathbf{h}_z$ using Eq. (\ref{eqn:HzMatrix}) finally leads to the following eigenvalue problem
\begin{equation}
\left[ \begin{matrix}
-\mathbf{k}_x \mathbf{k}_y & \mathbf{k}_x^2 - k_0^2 \bm{\varepsilon_y} \\
k_0^2 \bm{\varepsilon_x} - \mathbf{k}_y^2 & \mathbf{k}_y \mathbf{k}_x
\end{matrix} \right] \left[ \begin{matrix}
\mathbf{e}_x \\
\mathbf{e}_y
\end{matrix} \right] = \pm \omega \mu_0 \beta \left[ \begin{matrix}
\mathbf{h}_x \\
\mathbf{h}_y
\end{matrix} \right].
\label{eq:HxHyMat}
\end{equation}
The splitting of the factorization along the $x$ and $y$ axes such that the inverse rule can be used along the $x$ axis and the direct rule along the $y$ axis relies on the separability of the $k_x$ and $k_y$ dependencies of the discretization grid such that the discretization in \eqref{eq:etadisc} can be performed in a well-defined manner. However, for our dartboard discretization scheme, this separation is not possible, and for this reason, we simply use the direct rule for the factorization with $\bm{\varepsilon}_x=\bm{\varepsilon}_y=\bm{\varepsilon}_{\mathrm{Tot}}$. 

\section{Relationship between open and periodic boundary conditions} \label{app:pbc}

To understand the equivalence between the open BC formalism with equidistant discretization and the periodic BC formalism, let us consider the representation of a function $f(x)$ with compact support such that $f(x) = 0$ for $|x| > L/2$. The continuous integral expansion of this function is given by 
\begin{align}
	f(x) &= \int F(k) \exp (\ii k x) \dd k \label{eq_ex} \\
	F(k) &= \frac{1}{2\pi} \int_{-L/2}^{L/2} f(x) \exp (-\ii k x) \dd x, \label{eq_ex2}
\end{align}
where the integration domain in \eqref{eq_ex2} has been reduced from $[-\infty,\infty]$ to $[-L/2,L/2]$ since $f(x) = 0$ outside this range.

We now implement the equidistant discretisation scheme with a discretization step $\Delta k$ such that \eqref{eq_ex} becomes
\begin{align}
	f(x) &= \sum_n F(k_n) \exp (\ii k_n x) \Delta k, \label{eq_ex3}
\end{align}
where $k_n = n \Delta k$. 

Let us compare this equation to the Fourier series expansion of the same function over the interval $[-L/2,L/2]$ given by
\begin{align}
f(x) &= \sum_n c_n \exp (\ii k_n x) \label{eq_ex4} \\
c_n &= \frac{1}{L} \int_{-L/2}^{L/2} f(x) \exp (-\ii k_n x) \dd x. \label{eq_ex5}
\end{align}
where $k_n = n 2 \pi/L$. Now, the integral expansion (\ref{eq_ex})-(\ref{eq_ex2}) should ideally reproduce a function $f(x)$ for which $f(x) = 0$ for $|x| > L/2$. However, we observe that the representation in \eqref{eq_ex3} implementing the equidistant discretization is mathematically equivalent to the standard Fourier series representation (\ref{eq_ex4})-(\ref{eq_ex5}) of a periodic function $f(x)=f(x+L)$, where the periodicity is given by 
\begin{equation}
L = \frac{2\pi}{\Delta k}.
\end{equation}
When representing the optical fields using an open BC and equidistant discretization, we are thus in practice reintroducing a periodic BC with the associated numerical artifacts due to the presence of the neighboring elements. The artifacts can be suppressed by decreasing $\Delta k$, in which case the Riemann sum representation of the Fourier transform approaches the exact value of the integral. However, this occurs at the expense of significant computational cost, and a non-uniform discretization scheme is thus strongly preferred.

%******************************************
% Bibliography
%\bibliographystyle{osajnl}
%\bibliography{References_OpenFMM}
%\bibliography{C:/Fotonik/library}

\end{document}